\def\Cosh{{\rm Cosh}}
\def\Cos{{\rm Cos}}

\def\mpi{{\rm m}_\pi}
\def\Nc{{N_c}}
\def\sqr#1#2{{\vcenter{\vbox{\hrule height.#2pt
        \hbox{\vrule width.#2pt height#1pt \kern#1pt
           \vrule width.#2pt}
        \hrule height.#2pt}}}}
\def\square{\mathchoice\sqr84\sqr84\sqr53\sqr43}
\documentstyle[12pt]{article}
\begin{document}
\newcommand{\nd}[1]{/\hspace{-0.5em} #1}
\begin{titlepage}
\begin{flushright}
hep-ph/9509310\\
MSU-50605 \\
SWAT/85 \\
LA-UR-95/2991\\
September 1995  \\
\end{flushright}

\begin{centering}
\vspace{.6in}
{\Large {\bf Chiral Divergent Properties of Hadrons \\ in the Large
N$_{\bf c}$ Limit}}

\vspace{.5in}
{\bf Nicholas Dorey}$^1$, {\bf James Hughes}$^2$,
{\bf Michael Mattis}$^3$, and {\bf David Skidmore}$^1$ \\
\vspace{.2in}
$^1$Physics Department, University College of Swansea \\
Swansea SA2 8PP, UK \\
\vspace{.1in}
$^2$Physics Department, Michigan State University \\
East Lansing, MI 48823, USA \\
\vspace{.1in}
$^3$Theoretical Division, Los Alamos National Laboratory\\
Los Alamos, NM 87545, USA \\
\vspace{.9in}
{\bf Abstract} \\
\vspace{.05in}
\end{centering}
{\small We compute the leading non-analytic correction to the baryon
mass in the combined limits $m_{\pi}\rightarrow 0$,
$N_{c}\rightarrow\infty$ with $m_{\pi}N_{c}$ held fixed. We reproduce
the results of Cohen  and Broniowski
using semiclassical methods rather than heavy-baryon
chiral perturbation theory. The calculation is organized to demonstrate:
$\underline{1}$ the model independence of our results; and
$\underline{2}$ the crucial role played by (iso)rotations of
the hedgehog pion cloud surrounding the baryon.  Our
rotationally improved large-$\Nc$
calculation yields results that naturally interpolate between
ordinary chiral perturbation theory on the one hand, and
large-$N_c$ physics on the other.}
\end{titlepage}
\paragraph{}
Chiral perturbation theory and the large-$N_{c}$ expansion have both
proven to be useful approximation schemes for studying the dynamics of
mesons and baryons at low-energy. However it has long been suspected
that the limits $N_{c}\rightarrow \infty$ and $m_{\pi}\rightarrow 0$ do
not commute \cite{AN} and certainly there exist several hadronic
quantities for which neither approximation, by itself, is adequate. In
particular, this is always the case for quantities which receive a
significant contribution from the $\Delta$ resonance. If the limit
$N_{c}\rightarrow\infty$ is taken with $m_{\pi}$ fixed then the
$\Delta$ becomes a stable particle, degenerate with the nucleon. If,
on the other hand, one takes the limit $m_{\pi}\rightarrow 0$ with $N_{c}$
fixed then at sufficiently low momenta the $\Delta$
 decouples from low-energy
physics. We will refer to these alternative limits as the \it naive
\rm large-$N_{c}$ limit and \it
naive \rm chiral limit, respectively.
\paragraph{}
The reason for the failure of both expansions is best
understood in terms of the ratio $d$ between the $\Delta$-nucleon mass
splitting and the pion mass \cite{cohen}:
\begin{equation}
d\ =\ {M_\Delta-M_N\over m_{\pi}}\ \sim\
(N_{c}\,m_{\pi})^{-1}
\label{ddef}
\end{equation}
 In the real world $d\simeq 2.1$, which is
neither large enough to justify the chiral limit nor small enough for
the large-$N_{c}$ limit to be realistic. The quantities which are most
seriously afflicted by this problem are those which diverge in the
naive chiral limit. Here we will focus on one such quantity,
 $S=\partial^{2}M_N/\partial (m_{\pi}^{2})^{2}$.
In the naive chiral limit $S$ is dominated by the
leading non-analytic correction to the baryon mass, $\delta M_N\sim
m_{\pi}^{3}$, hence $S\sim m_{\pi}^{-1}$.
Other chiral divergent quantities to which these remarks apply
include the isovector electric
and magnetic charge radii of the nucleon, $\langle r^{2}
\rangle^{E,M}_{I=1}$
and the isoscalar electric polarizability, $\alpha=(\alpha_{p}+\alpha_{n})/2$.
\paragraph{}
Recently it has been suggested that for a realistic
treatment of the $\Delta$ it is advisable formally to link the two limits,
by taking $m_{\pi}\sim N_{c}^{-1}$.
This alternative expansion scheme has been proposed both
in the context of the effective Lagrangian for baryons \cite{cohen} and
(by the present authors)
 in the Skyrme model \cite{dhmI}, and is also implicit in the heavy-baryon
chiral perturbation theory as developed by
Jenkins and Manohar \cite{jenkins}. In the former approach \cite{cohen},
by calculating Feynman diagrams in chiral perturbation theory, Cohen
and Broniowski have
obtained each of the quantities mentioned above as  functions of the
variable $d$, which reduce to the naive chiral and large-$N_{c}$
results in the limits $d\rightarrow \infty$ and $d\rightarrow 0$
respectively. The purpose of this letter is to argue that,
contrary to recent claims \cite{cohenII},
the correct behaviour for $d\neq 0$ is
equally obtainable in the Skyrme model or, more generally, in \it any \rm
semiclassical
hedgehog model of large-$N_{c}$ baryons.
\paragraph{}
 In the Skyrme model, the leading
order in the new expansion is obtained by considering a new field
configuration, the Rotationally Improved Skyrmion (RISKY). The RISKY
differs from the ordinary Skyrmion by including the deviation from
hedgehog structure which is induced by the collective coordinate
rotation of the soliton. Given that the $\Delta$ resonance in the
Skyrme model is a rotational excitation of the soliton with
$M_\Delta- M_N=3/2\Lambda$, where $\Lambda\sim N_{c}$ is the
classical moment of inertia, it is easy to show that the rotational
corrections to the large-distance asymptotics of the
Skyrmion have precisely the form of a power series in $d$.
In Ref.~\cite{dhmI}, we showed how the RISKY allows
a physically accurate description of $\Delta$-decay in the Skyrme
model. In the following we will demonstrate how essentially the same
effect provides the correct behaviour for the quantities
mentioned above.
\paragraph{}
The class of models for which the pion field has a
hedgehog structures includes not only the Skyrme model,  but also
the traditional chiral bag model, and the semiclassical meson-baryon
``large-$N_c$ bag models''
recently introduced by the present authors \cite{dhmII,dm} and independently
by Manohar \cite{manohar} (and  applied to the $NN$ problem
by Schwesinger \cite{schwes}). In the following we will concentrate on
the latter type of models. In particular, Manohar has shown
 how a semiclassical
effective theory correctly reproduces the leading $O(m_\pi^3)$
non-analytic
correction to the baryon mass in the naive large-$N_c$ limit.
However, as calculated below, the leading non-analytic
behaviour in the new expansion actually goes like $m_{\pi}^{3}$ multiplied by
some non-trivial function of $d$, so that Manohar's calculation is strictly
valid only for $d\rightarrow 0$. This is because Manohar
explicitly neglects effects due to the collective coordinate rotation
of the hedgehog baryon.
\paragraph{}
The main results of this letter are twofold. First, the $d$-dependent
effects \it can \rm be systematically included,
and the correct dependence on $d$ obtained, purely within the framework of
semiclassical physics. Thus, semiclassical methods provide a useful
alternative to diagram-by-diagram chiral perturbation theory.
Second, since chirally
divergent quantities in any hedgehog model are dominated by the
contribution of the large distance pion tail, and since this tail has a
universal form, the results obtained below are
pleasingly \it model independent\rm.
\paragraph{}
Let us consider an effective Lagrangian for pions coupled to
non-relativistic large-$N_{c}$ baryons as described in
\cite{dhmII,dm,manohar}.
The purely pionic part of the Lagrangian is given by
\begin{equation}
{\cal L}_{\pi}\ = \ \frac{f_{\pi}^{2}}{16}{\rm
Tr}\,\partial_{\mu}U
\partial^{\mu}U^{\dagger}\ +\
\frac{m_{\pi}^{2}f_{\pi}^{2}}{8}{\rm Tr}\left(U-1\right)\ +\ \cdots
\label{lpi}
\end{equation}
where $U=\exp(2i{\vec \pi}\cdot{\vec \tau}/f_{\pi})\in SU(2)$ and
the dots denote higher derivative terms in the chiral
expansion. In addition, the pions are coupled to a tower of \it explicit \rm
non-relativistic baryons with $I=J=1/2,3/2,\ldots$ as dictated by
(2-flavor)
large-$N_{c}$ selection rules. The baryon tower can be compactly described in
terms of an $SU(2)$ collective coordinate $A(t)$ defined by the change
of basis
\begin{equation}
\langle I=J, i_{z}, s_{z}|A\rangle = ({2J+1})^{1/2}D_{-s_{z},i_{z}}^{(J)}
(A^\dagger)\cdot(-)^{J-s_{z}}
\label{wigd}
\end{equation}
where $D^{(J)}(A)$ is a Wigner representation matrix of spin/isospin $J$. In
the $|A\rangle$ basis the pion-baryon coupling is diagonal:
\begin{equation}
{\cal L}_{\pi B}\ = \ \frac{3g^{b}_{\pi}}{2M}\delta^{(3)}({\bf
x})D^{(1)}_{ia}(A(t)){\cal J}^{A}_{ia}[U]
\end{equation}
where ${\cal J}^A$ is the pion axial current and $g^{b}_{\pi}$ is the
bare pion-nucleon coupling constant.
The rotational
effects which generate the $\Delta$-nucleon mass splitting precisely
correspond to the {\em time-dependence} of the matrix $A(t)$. This
time dependence is governed by the
Lagrangian for free motion on the $SU(2)$ group manifold:
\begin{equation}
L_{B}=-M+\Lambda{\rm Tr}\,\dot{A}^{\dagger}\dot{A}
\label{alag}
\end{equation}
where $M$ is the baryon mass and $\Lambda\sim N_{c}$
is the moment of inertia for this rotation.\footnote{As explained in
detail in \cite{dhmII,dm,manohar},
the bare parameters $g^b_\pi,$ $M$ and $\Lambda$
are actually renormalized by an infinite class of UV divergent
Feynman diagrams which contribute at leading order
in the $1/N_{c}$ expansion. It is therefore essential to regulate
the theory, which is most easily accomplished by
smearing out the $\delta$-function source over a radius $R$ in some
way: $\delta^{(3)}({\bf x})  \rightarrow  \delta^{(3)}_{R}({\bf x}).$
For simplicity we gloss over most of these interesting
complications in the following discussion.} Quantizing this degree of
freedom gives rise to the $\Delta$-nucleon mass splitting.
The large moment of
inertia means that isorotation is suppressed as
$N_{c}\rightarrow\infty$ and hence this effect is neglected in the
calculation of Manohar.
\paragraph{}
The main result of \cite{dhmII,dm,manohar}
is that this model can be solved exactly
in the large-$N_{c}$ limit by solving an appropriate Euler-Lagrange
equation. Further, for $|{\bf x}|$ much greater than the scale of
chiral symmetry breaking $\Lambda_{\chi}$, the magnitude of
the pion field is small and the equation becomes linear:
\begin{equation}
(\square+m_{\pi}^{2})\pi^a = \frac{3g^{r}_{\pi}}{2M}
D_{ai}^{(1)}\big(A(t)\big)\frac{\partial}{\partial x^{i}}
\delta^{(3)}_{R}({\bf x})
\label{lceq}
\end{equation}
Here $g^{r}_{\pi}$ is the renormalized pion-nucleon
coupling constant which is a
complicated function of the cutoff $R$ and the bare coupling $g^{b}_{\pi}$.
In \cite{dm} we showed how these two parameters can be varied
simultaneously so that $g^{r}_{\pi}$ remains constant. This defines a
renormalization group (RG) flow for the model. In particular, if the Skyrme
term is included in the Lagrangian (\ref{lpi}), it can be shown that the Skyrme
model itself arises as the UV fixed point of this RG flow: the simultaneous
limit $R\rightarrow 0$, $g^{b}_{\pi}\rightarrow 0$.
For this reason we stress that the Skyrme model should be thought of
as a particular member of the general class of models considered here.
\paragraph{}
The advantage of equation (\ref{lceq})
 is that, because of its
linearity, it can be solved immediately by the method of Green's
functions:
\begin{equation}
\pi^a({\bf x},t) = -\frac{3g_{\pi}^{r}}{2M}
\int d^4x' G_F(x-x')D_{ai}^{(1)}\big(A(t')\big)
\frac{\partial}{\partial x'_{i}}
\delta^{(3)}_{R}({\bf x}')
\label{lsol}
\end{equation}
where the pion Green's function $G_F$ which
satisfies Feynman boundary conditions is given by
\begin{equation}
G_F(x-x') = \int \frac{d^4k}{(2\pi )^4}
\frac{e^{ik\cdot (x-x')}}{k^2 - m_{\pi}^2 + i\epsilon}
\label{gfct}
\end{equation}
\paragraph{}
In order to make progress with the expression (\ref{lsol}) it is
necessary to know the time-dependence of the $SU(2)$ collective
coordinate $A(t)$. The first and simplest case is the
naive large-$N_{c}$ limit with $m_{\pi}$ held fixed. Here the
moment of inertia becomes infinite and the collective coordinate is
frozen; $A(t)\equiv A$. In this case the integrals may be performed
explicitly and the resulting pion cloud has the hedgehog form;
$\pi^{a}({\bf x})=\textstyle{1\over2}f_\pi F(r)D_{ai}^{(1)}(A)\hat{x}^i$
where $r=|{\bf x}|$ and $\hat{x}^i=x^i/r.$ Asymptotically, the hedgehog
profile approaches
\begin{equation}
F(r)\rightarrow B\cdot\left(\frac{m_{\pi}}{r}+\frac{1}{r^{2}}\right)
\exp\left(-m_{\pi}r\right)
\label{asymptotics}
\end{equation}
where $B=3g^{r}_{\pi}/4\pi f_\pi M$.
Hence the static pion cloud has exactly the same
far-field behaviour as the Skyrmion. In our case, although the moment
of inertia $\Lambda\sim N_{c}$, we also have $m_{\pi}\sim
N_{c}^{-1}$, hence rotational corrections which depend on these
parameters in the combination $\Lambda m_{\pi}$ must be retained. As
stated above, the time evolution of $A(t)$ is governed by the
Hamiltonian obtained from (\ref{alag}).
Hence in order to obtain the correct result
we should promote the classical collective coordinate $A$ in
expression (\ref{lsol}) to a quantum
operator $\hat{A}$ and take an
expectation value with respect to this Hamiltonian in a particular
baryon state. In the following we
will use exactly this approach to calculate the  ``rotational
improvement'' of Manohar's result for the leading non-analytic
contribution to the baryon mass. However, before describing this
calculation, it is instructive to consider the modified field
configuration which includes the relevant rotational corrections.
\paragraph{}
As is well known, the
quantum mechanics of a free particle moving on the $SU(2)$ group
manifold is semiclassically exact \cite{schulman}. As a consequence,
quantum expectation values can be replaced by summations
over {\em classical} paths of fixed (classical) angular momentum
${\bf J}$. Such a path corresponds to geodesic motion on the $SU(2)$
group manifold; $A(t)=\exp(i{\bf \omega}t\cdot{\vec\tau}/2)A(0)$, where
${\bf \omega}=|{\bf J}|/\Lambda$. Hence the relevant pion field
configuration is obtained by substituting this path for $A(t)$ in
(\ref{lsol}). This is straightforward and yields $\pi_{a}({\bf x},t)=
D_{ai}^{(1)}(A(t))\tilde{\pi}_{i}({\bf x};{\bf J})$
where,
\begin{eqnarray}
\tilde{\bf \pi}({\bf x},{\bf J}) & \rightarrow & \frac{f_\pi B}
{2{\bf J}^{2}}\Bigg\{  \left(
\frac{m_{\pi}}{r}+\frac{1}{r^{2}}\right)\exp\left(-m_{\pi}r\right)\left({\bf
J}\cdot\hat{\bf r}\right){\bf J}  \nonumber \\
&  & \quad{}  +
\left(\frac{\sqrt{m^{2}_{\pi}-{\bf J}^{2}/\Lambda^2}}{r}+\frac{1}{r^{2}}\right)
\exp\left(-\sqrt{m^{2}_{\pi}-{\bf J}^{2}/\Lambda^2}\cdot r\right)\left({\bf
J}\times {\bf J} \times \hat{\bf r}\right) \ \Bigg\}\ .\nonumber \\
\label{mf}
\end{eqnarray}
This is precisely the asymptotic behaviour of the RISKY which we
obtained by different means in the Skyrme model in
\cite{dhmI}. Clearly the difference between this asymptotic behaviour
and that of the standard Skyrmion can be expressed as a power series
in $(\Lambda m_{\pi})^{-1}$ or, equivalently, a power series in $d$.
In practise, calculating the sum over such field configurations with
different values of ${\bf J}$ to obtain, for
example, the pion one-point Green's function is somewhat subtle.
In the analysis of the Skyrme model given in \cite{dhmI}, the RISKY was
quantized directly by introducing the operators
$\hat{A}$ and $\hat{\bf J}$. However as these are non-commuting operators, it
was still necessary to resolve the consequent operator ordering ambiguity by
appealing to physical principles. In the present case, because we have
the effective linear field equation (\ref{lceq}),
all of these difficulties can
be avoided by using the Green's function solution (\ref{lsol}).
\paragraph{}
{\bf The Calculation:}
As a concrete illustration of the ideas presented above,
let us calculate  the  contribution $\delta M_N$ to
the nucleon mass due
to the far field of the strongly coupled pion cloud.
At leading order this is given by the linearized
Lagrangian whose variation gives the field equation (\ref{lceq}):
\begin{equation}
2T\,\delta M_N \ = \ \langle N,T\vert\, \int_{-T}^{T}d^4x \,\pi^a_{\rm cl}
({\bf x}, t)\Big(\,{\textstyle{1\over2}}
(\square+m_{\pi}^{2})\pi_{\rm cl}^{a}({\bf x}, t)
-
 \frac{3g_{\pi}}{2M_N}
D_{ai}^{(1)}\big(A(t)\big)\frac{\partial}{\partial x^{i}}
\delta^{(3)}({\bf x})\,\Big)\,
\vert N, -T\rangle
\label{delm}
\end{equation}
As with the evaluation of the Yukawa coupling \cite{dhmI}
we begin by inserting the far field pion given in Eq.~(\ref{lsol})
into the right hand side of Eq.~(\ref{delm}) to get
\begin{eqnarray}
2T\,\delta M_N  &
= & \frac{9g_{\pi}^2}{8M_N^2}
\int d^4x\,d^4x'\int\frac{d^4k}{(2\pi )^4}
\frac{e^{ik\cdot(x-x')}}{k^2 - m^2 + i\epsilon}\,
\frac{\partial}{\partial x^{}_{l}}
\delta^{(3)}({\bf x})
\frac{\partial}{\partial x'_{m}}
\delta^{(3)}({\bf x}')
\nonumber \\ &  & \qquad \qquad \times
\langle N, T \vert {\cal T} \{ D_{al}^{(1)}\big(A(t)\big)
D_{am}^{(1)}\big(A(t')\big)\big\}  \vert N, -T \rangle
\nonumber\\
& = & -\frac{3ig_\pi^2}{8M_N^2}
\int_{-T}^T dt\,dt'\int\frac{d^3{\bf k}}{(2\pi)^3\, 2\omega_{\bf k}}\,
{\bf k}^2\,e^{-i\omega_{\scriptscriptstyle\bf k}|t-t'|}
\nonumber \\ &  & \qquad \qquad \times
\langle N, T \vert {\cal T} \{ D_{am}^{(1)}\big(A(t)\big)
D_{am}^{(1)}\big(A(t')\big)\big\}  \vert N, -T \rangle
\nonumber\\
\label{delmone}
\end{eqnarray}
In obtaining the second equality we have pulled the derivatives
off the $\delta$-functions by integrating by parts, replaced
$k^lk^m$ by ${1\over3}\delta^{lm}{\bf k}^2$ by spherical symmetry,
and carried out the $k_0$ integration.
The expression for the far field as a functional of $A(t)$ together with
the path integral implicit in Eq.~(\ref{delmone}) defines a
quantum mechanical problem with operator
$A(t) \in SU(2)$. In particular
the time-ordered quantum mechanical two-point function
in (\ref{delmone}) may be resolved by inserting a complete set of intermediate
nucleon and $\Delta$ states $|I\rangle\equiv|I,i_z,s_z\rangle$:
\begin{eqnarray}
& &\langle N, T \vert {\cal T} \{
D_{am}^{(1)}\big(A(t)\big)
D_{am}^{(1)}\big(A(t')\big)\big\} \vert N, -T \rangle
\nonumber \\ & &
=\ e^{2iM_NT}\sum_{I={1\over2},{3\over2}}\,\sum_{i_z,s_z=-I}^I
\ \theta(t'-t)\,e^{-iM_N(T-t')}\langle N |
D_{am}^{(1)}(A)
| I\rangle
e^{-iM_I(t'-t)}
\langle I |
D_{am}^{(1)}(A)
| N\rangle
e^{-iM_N(t+T)}\nonumber
\\ & & \qquad+\ \ (t\leftrightarrow t')\ ,
\label{top}
\end{eqnarray}
where the exponential prefactor represents the need to divide out
by  free nucleon propagation, $\langle N,T|N,-T\rangle.$
As we shall invoke in a moment, the remaining single-particle matrix elements
$\langle I |D_{am}^{(1)}(A)| N\rangle$
are easily evaluated using the wave
functions $\langle N \vert A \rangle $ given in Eq.~(\ref{wigd}).

Inserting Eq.~(\ref{top})
into Eq.~(\ref{delmone}), passing to the $T\rightarrow\infty$
limit, and performing the time integrations gives
\begin{eqnarray}
& &\delta M_N\ =\
-\frac{3g_\pi^2 }{4M_N^2}
\sum_{I={1\over2},{3\over2}}\,\sum_{i_z,s_z=-I}^I
{\langle N | D_{am}^{(1)}(A) | I\rangle}^2
\int\frac{d^3{\bf k}}{(2\pi)^3}\frac{{\bf k}^2}{2\omega_{\bf k}(\omega_{\bf k}
+M_I-M_N)}
\end{eqnarray}
Note that the remaining integral
 is divergent in the ultraviolet. Technically, this is because, in this
section, we used an exact $\delta$-function $\delta^{(3)}({\bf x})$
rather than a smeared-out source $\delta^{(3)}_R({\bf x})$ such as in
Eq.~(\ref{lceq}); on a more fundamental level, discussed in great detail in
Refs.~\cite{dhmII,dm}, this infinity is a reflection of the UV divergences in
the underlying pion-baryon Feynman diagrams that are summed implicitly
by our semiclassical methods. However, in the present paper, it is
the \it infrared \rm properties of the theory that we are interested in,
in particular the nonanalytic $m_\pi^3$ contribution to $\delta M_N.$
This quantity is finite, and is most easily accessed by differentiating
twice with respect to $m_\pi^2$.
 One then obtains, when the intermediate
baryon $I$ is a $\Delta$, the convergent result
\begin{equation}
{\partial^2\over \partial(m_\pi^2)^2}\,
\int {d^3{\bf k}\over(2\pi)^3}{{\bf k}^2\over 2\omega_{\bf k}
(\omega_{\bf k}+M_I-M_N)}\ =\ {3\over16\pi^2m_\pi}{\Cosh^{-1}d
\over\sqrt{d^2-1}}
\end{equation}
where we assume $d>1$ as in Nature; if instead $I$ is a nucleon
the right-hand side is simply $\big(\textstyle{32\over3}\pi m_\pi\big)^{-1}.$
Furthermore, the Clebsch-Gordan expression $\sum_{i_z,s_z}
\langle N | D_{am}^{(1)}(A) | I\rangle^2$ works out to
2 and 1, respectively, for the two cases $I=\Delta$ and $I=N.$
So the final answer, summed over both intermediate nucleons and
$\Delta$'s, reads:
\begin{equation}
{\partial^2\over\partial(m_\pi^2)^2}
\delta M_N\ =\
-{9g_\pi^2\over128 M_N^2\pi m_\pi}\Big(\,1+{4\over\pi}{\Cosh^{-1}d
\over\sqrt{d^2-1}}\,\Big)\ ,\qquad d>1\ .
\end{equation}

This result is identical to that found by Cohen and Broniowski in \cite{cohen}.
It correctly interpolates between the result
from naive chiral calculations where $d \rightarrow \infty$, and the
three times larger naive large-$\Nc$ result \cite{manohar}, where
$d \rightarrow 0$; of course, in the latter limit, one needs to analytically
continue the above expression around the branch point,
obtaining instead
\begin{equation}
{\partial^2\over\partial(m_\pi^2)^2}
\delta M_N\ =\
-{9g_\pi^2\over128 M_N^2\pi m_\pi}\Big(\,1+{4\over\pi}{\Cos^{-1}d
\over\sqrt{1-d^2}}\,\Big)\ ,\qquad d<1\ .
\end{equation}
\paragraph{}
In summary, naive $\Nc \rightarrow \infty$ and $\mpi \rightarrow 0$ limits
do not commute.  Yet for certain important quantities, both limits,
as well as the interpolation between them,
may be extracted from a single common semiclassical starting point.
Furthermore, the calculation presented above is model independent,
 encompassing both large-$N_c$  bag models and
skyrmion models by virtue of the large-$\Nc$ RG,

JH is indebted to M. Rho and the Institute for Nuclear Theory,
Seattle, for hospitality during the initial stages of this
work, and to T. Cohen for valuable discussions.

\end{document}